\documentclass[prc,twocolumn,showpacs,preprintnumbers,amsmath,amssymb,superscriptaddress,floatfix,nofootinbib]{revtex4}
\usepackage{graphicx}
\usepackage{amsmath}
\usepackage{amsfonts}
\usepackage{amssymb}
\usepackage{color}

\begin{document}

\title{On the coupling constant for $\Lambda(1405)\bar{K}N$}
\author{Ju-Jun Xie} \email{xiejujun@impcas.ac.cn}
\affiliation{Institute of Modern Physics, Chinese Academy of Sciences, Lanzhou 730000, China}
\affiliation{State Key Laboratory of Theoretical Physics, Institute of Theoretical Physics, Chinese Academy of Sciences, Beijing 100190,
China}

\author{Bo-Chao Liu} \email{liubc@xjtu.edu.cn} \affiliation{Department of Applied Physics, Xi'an
Jiaotong University, Xi'an, Shanxi 710049, China}
\affiliation{State Key Laboratory of Theoretical Physics, Institute of Theoretical Physics, Chinese Academy of Sciences, Beijing 100190,
China}

\author{Chun-Sheng An} \email{ancs@ihep.ac.cn}
\affiliation{Institute of High Energy Physics, and Theoretical Physics Center for Science Facilities,
Chinese Academy of Sciences, Beijing 100049, China}

\begin{abstract}

The value of $\Lambda(1405){\bar K}N$ coupling constant
$g_{\Lambda(1405)\bar{K}N}$ is obtained by fitting it to the
experimental data on the total cross sections of the $K^-p \to
\pi^0\Sigma^0$ reaction. On the basis of an effective Lagrangian
approach and isobar model, we show that the value
$|g_{\Lambda(1405)\bar{K}N}| = 1.51 \pm 0.10$ could be extracted
from the available experimental data by assuming that the
$s-$channel $\Lambda(1405)$ resonance plays the dominant role, while
the background contributions from the $s-$channel $\Lambda(1115)$,
$t-$channel $K^*$ and $u-$channel nucleon pole processes are small
and can be neglected. However, the $u-$channel nucleon pole diagram
may also give an important contribution in present calculations.
After the background contributions are taken into account, the above
value of $g_{\Lambda(1405)\bar{K}N}$ is reduced to
$|g_{\Lambda(1405)\bar{K}N}| = 0.77 \pm 0.07$, which is not
supported by the previous calculations and the recent CLAS
measurements. The theoretical calculations on differential cross
sections are also presented, which can be checked by the future
experiments.

\end{abstract}
\maketitle

\section{Introduction}

The reactions induced by $K^-$ meson beam are important tools to
gain a deeper understanding of the $\bar{K}N$ interaction and also
of the nature of the hyperon resonances. Among those reactions, the
$K^- p \to \pi^0 \Sigma^0$ reaction is of particular interest. Since
there are no isospin-1 hyperons contributing here, this reaction
gives us a rather clean platform to study the isospin zero $\Lambda$
resonances. Furthermore, it is well known that the inelastic effects
are especially important for the low energy $\bar{K}N$ interaction
because the $\bar{K}N$ channel strongly couples to the $\pi\Sigma$
channel through $\Lambda(1405)$ resonance (spin-parity $J^P =
1/2^-$). Thus, the $K^- p \to \pi^0 \Sigma^0$ reaction is a good
place to study the $\Lambda(1405)$ state, whose structure and
properties are still controversial, even it is catalogued as a
four-star $\Lambda$ resonance in the Particle Data Group (PDG)
review book~\cite{pdg2012}.

In the traditional quark models, the $\Lambda(1405)$ is described as
a $p-$wave $q^3$ baryon~\cite{isgurprd18}, but it can also be
explained as a $\bar{K}N$ molecule~\cite{dalitz} or $q^4\bar{q}$
pentaquark state~\cite{inouenpa790}. Besides, it was also argued,
within the unitary chiral
theory~\cite{kaisernpa612,garciaplb582,chiraloset}, two overlapping
isospin $I=0$ states are dynamically generated and in this approach
the shape of any observed $\Lambda(1405)$ spectrum might depend upon
the production process. In a recent experimental study of the $pp\to
pK^+\Lambda(1405)\to pK^+ (\pi\Sigma)$ reaction~\cite{zychorplb660},
the $\Lambda(1405)$ resonance was clearly identified through its
$\pi^0\Sigma^0$ decay and no obvious mass shift was found, which has
been checked in Ref.~\cite{Xie:2010md} by using the effective
Lagrangian approach with considering only one $\Lambda(1405)$ state.
However, the final answer is still absent in the sense that the
experimental data could also be well described in the two-resonance
scenario~\cite{gengepja34}.

As shown in
Refs~\cite{Tsushima:2000hs,Sibirtsev:2005mv,Shyam:1999nm,liuplk,Xie:2007vs,xiephi,gaopzkn},
the combination of effective Lagrangian approach and the isobar
model is a good method to study the hadron resonances production in
the $\pi N$, $NN$, and $\bar{K}N$ scattering. One key issue of this
method is the coupling constant of the involved resonance
interaction vertex, which can be obtained from the partial decay
width. However, if the mass of the resonance is below the threshold
of the corresponding channel, such as the $\Lambda(1405)$ ($M=1405$
MeV) to the $\bar{K}N$ ($m_{\bar{K}}+m_N=1434.6$ MeV) channel, it is
impossible to get the coupling constant within the above procedure.
For example, the strong coupling constants of $\Lambda(1405)$
resonance were investigated within an extended chiral constituent
quark model~\cite{An:2010wb}, while in
Refs.~\cite{xiephi,Dai:2011yr}, the coupling constant of
$g_{N(1535)N\phi}$ was obtained from the studies of the $\pi^- p \to
n\phi$ reaction. Besides, in Ref.~\cite{Xie:2008ts}, the coupling
constant of $g_{N^*(1535)N \rho}$ was studied from the analysis of
the $N^*(1535) \to N \rho^0 \to N\pi^+ \pi^-$ and the $N^*(1535) \to
N\rho^0 \to N\gamma$ decays.

Moreover, the couplings of $\Lambda(1405)$ resonance to the
$\bar{K}N$ and $\pi\Sigma$ channels and the ratio of
$g_{\Lambda(1405)\bar{K}N}$ and $g_{\Lambda(1405)\pi\Sigma}$,
$R=g_{\Lambda(1405)\bar{K}N}/g_{\Lambda(1405)\pi\Sigma}$, have been
intensively studied both experimentally~\cite{Exp} and within
various theoretical approaches, for instance, the current
algebra~\cite{Weil,G-M}, potential models~\cite{Weil,Dalitz},
dispersion relations~\cite{Martin}, asymptotic $\text {SU}(3)$
symmetry approach~\cite{Oneda}, and they are also recently
investigated by taking the $\Lambda(1405)$ resonance to be an
admixture of traditional three-quark and higher five-quark Fock
components~\cite{An:2010wb}. However, the obtained values of $R$ by
different theoretical methods vary in a large range $3.2-7.8$, so it
is still worth to study the coupling constants
$g_{\Lambda(1405)\bar{K}N}$ and $g_{\Lambda(1405)\pi\Sigma}$ from
other different ways.

The present work we report here is one more in this line, by using
an effective Lagrangian approach and the isobar model, we extract
the $\Lambda(1405){\bar K}N$ coupling constant
$g_{\Lambda(1405)\bar{K}N}$ by fitting it to the experimental data
on the total cross sections of $K^- p \to \pi^0 \Sigma^0$ reaction
near threshold. We also calculate the differential cross sections
for the $K^- p \to \pi^0 \Sigma^0$ reaction with the fitted
parameters. These model predictions can be checked by the future
experiments. For simplicity we shall here work within the single
$\Lambda(1405)$ state framework with parameters as reported in the
PDG~\cite{pdg2012}.

This article is organized as follows. In Sect.~\ref{Formalism}, we
present the formalism and ingredients necessary for our
calculations. We present in Sect.~\ref{Results} the values of the
obtained coupling constants. The theoretical calculations on the
differential cross sections are also shown. A short summary is given
in the last section.

\section{Formalism and ingredients} \label{Formalism}

The combination of  isobar model and effective Lagrangian method is
a useful theoretical approach in the description of various
processes in the resonance production region. In this section, we
introduce the theoretical formalism and ingredients for calculating
the cross sections of the $K^-p \to \pi^0\Sigma^0$ reaction within
the effective Lagrangian method.

\begin{figure}[htbp]
\begin{center}
\includegraphics[scale=0.61]{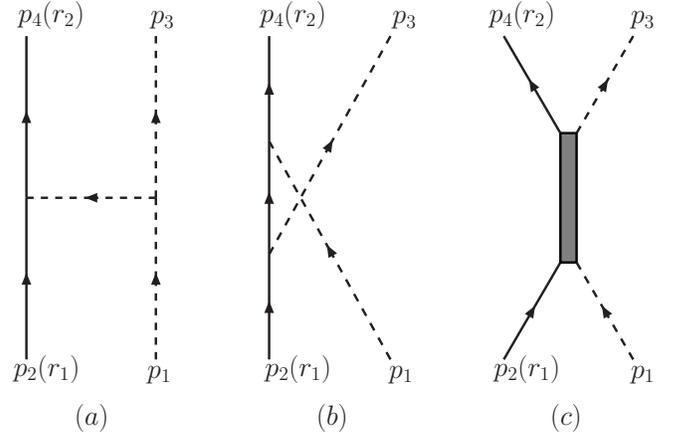}
\caption{Feynman diagrams for the reaction $K^- p \to
\pi^0\Sigma^0$. In these diagrams, we show the definitions of the
kinematical ($p_1, p_2, p_3, p_4$) and polarization variables $r_1,
r_2$ those we use in our calculation.} \label{feynfig}
\end{center}
\end{figure}

The basic tree level Feynman diagrams, for the $K^-p \to
\pi^0\Sigma^0$ reaction, are shown in Fig.~\ref{feynfig}. These
include the $t-$channel $K^*$ exchange [Fig.~\ref{feynfig} (a)], the
$u-$channel proton exchange [Fig.~\ref{feynfig} (b)], and the
$s-$channel $\Lambda(1115)$ and $\Lambda(1405) (\equiv\Lambda^*)$
processes [Fig.~\ref{feynfig} (c)]. To compute the contributions of
these terms, we use the effective interaction Lagrangian densities
as used in Refs.~\cite{wufq,Zou:2007rx,Mosel,feuster,Doring:2010ap}:
\begin{eqnarray}
{\cal L}_{K^* K \pi} &=& -g_{K^* K\pi}(\vec{\pi}\cdot
\tau\partial^\mu{\bar{K}}-\bar{K}\partial^\mu{\vec{\pi}}\cdot\tau)K^{*}_\mu
, \label{ksnl} \\
{\cal L}_{K^*N\Sigma}&=&-i g_{K^*N\Sigma}\bar{N}\big(\gamma_\mu
-{\kappa \over 2M_N}\sigma_{\mu\nu}\partial^\nu \big)
K^{*\mu}\vec{\Sigma}\cdot\tau \nonumber
\\ && + \mathrm{H.c.} \,,
\end{eqnarray}
for the $t-$channel $K^*$ exchange, and
\begin{eqnarray}
{\cal L}_{\pi NN}&=& -ig_{\pi NN}\bar{N}\gamma_5 \vec{\pi} \cdot\tau
N+\mathrm{H.c.}, \\
{\cal L}_{KN\Sigma}&=& -ig_{KN\Sigma}\bar{N}\gamma_5\vec{\Sigma}
\cdot \tau K+\mathrm{H.c.}, \\
{\cal L}_{KN\Lambda}&=& -ig_{KN\Lambda}\bar{N}\gamma_5\Lambda
K+\mathrm{H.c.},
\end{eqnarray}
for $u-$channel proton pole diagram, while
\begin{eqnarray}
{\cal L}_{\pi\Sigma\Lambda}&=&
-ig_{\pi\Sigma\Lambda}\bar{N}\gamma_5\Lambda
\vec{\pi}\cdot\vec{\Sigma}+\mathrm{H.c.}, \\
{\cal L}_{\Lambda^*{\bar K}N}&=& -i g_{\Lambda^*
{\bar K}N} \bar\Lambda^* \bar{K} N+\mathrm{H.c.}, \\
{\cal L}_{\Lambda^* \pi\Sigma}&=& -i g_{\Lambda^* \pi\Sigma}
\bar{\Lambda}^* \vec{\pi}\cdot \vec{\Sigma} +\mathrm{H.c.},
\end{eqnarray}
for the $s-$channel $\Lambda(1115)$ and $\Lambda(1405)$ terms.

For the coupling constants in the above Lagrangian densities, we
take $g_{\pi NN} = 13.45$ (obtained with $g^2_{\pi NN}/4\pi =
14.4$), $g_{KN\Lambda} = -13.98$, $g_{\pi \Sigma \Lambda} = 9.32$,
and $g_{KN\Sigma} = 2.69$ as that determined within SU(3) flavor
symmetry~\cite{swart6365}~\footnote{We show here the relations of
these coupling constants in terms of $g_{\pi NN}$ and $\alpha =
0.4$: $g_{KN\Lambda} = -\frac{g_{\pi NN}}{\sqrt{3}} (1+2\alpha)$,~~
$g_{KN\Sigma} = g_{\pi NN}(1-2\alpha)$, and $g_{\pi\Sigma\Lambda} =
\frac{2g_{\pi NN}}{\sqrt{3}} (1-\alpha)$.}. For the $K^*N\Sigma$
couplings, we take $g_{K^*N\Sigma} = -2.36$ and $\kappa = -0.47$ as
used in Ref.~\cite{Kim:2011rm} for the calculation of $K^*\Lambda$
photoproduction. While the coupling constants $g_{K^*K\pi}$ and
$g_{\Lambda^* \pi \Sigma}$ are determined from the experimentally
observed partial decay widths of the $K^* \to K\pi$ and
$\Lambda(1405) \to \pi \Sigma$, respectively,
\begin{eqnarray}
\Gamma_{K^* \to K \pi} &=& \frac{g^2_{K^*K\pi}}{2\pi
}\frac{|\vec{p_{\pi}}^{{\rm c.m.}}|^3}{m^2_{K^*}}, \label{kstarkpi} \\
 \Gamma_{\Lambda^* \to \pi \Sigma} &=& \frac{3
g^2_{\Lambda^*\pi \Sigma}}{4\pi}
(E_{\Sigma}+m_{\Sigma})\frac{|\vec{p}_{\Sigma}|}{M_{\Lambda^*}},
\end{eqnarray}
where
\begin{eqnarray}
|\vec{p_{\pi}}^{{\rm c.m.}}| &=& \frac{\sqrt{[m^2_{K^*}-(m_K+m_{\pi})^2][m^2_{K^*}-(m_K-m_{\pi})^2]}}{2m_{K^*}}, \nonumber \label{pcmpi} \\
E_{\Sigma} &=& \frac{M^2_{\Lambda^*}+m^2_{\Sigma}-m^2_{\pi}}{2M_{\Lambda^*}}, \nonumber \\
|\vec{p}_{\Sigma}| &=& \sqrt{E^2_{\Sigma}-m^2_{\Sigma}} \nonumber \,
.
\end{eqnarray}

With masses ($m_{K^*} = 893.1$ MeV, $m_{K} = 495.6$ MeV, $m_{\pi} =
138.04$ MeV, and $M_{\Lambda^*} = 1405.1^{+1.3}_{-1.0}$ MeV), total
decay widths ($\Gamma_{K^*} = 49.3$ MeV and $\Gamma_{\Lambda^*} = 50
\pm 2$ MeV), and decay branching ratio of $K^* \to K\pi$ and
$\Lambda(1405) \to \pi \Sigma$ [Br($K^* \to K\pi$) $\sim 1$ and
Br($\Lambda(1405) \to \pi \Sigma$) $\sim 1$], we obtain $g_{K^*K\pi}
= 3.25$, and $|g_{\Lambda^* \pi \Sigma}| = 0.90 \pm 0.02$ by
considering the uncertainties of the total decay width and the mass
of $\Lambda(1405)$ resonance.

With the effective Lagrangian densities given above, we can easily
construct the invariant scattering amplitude,
\begin{eqnarray}
{\cal M}_i=\bar u_{r_2}(p_4)~{\cal A}_i~u_{r_1}(p_2),
\end{eqnarray}
where $i$ denotes the $s-$, $t-$ or $u-$channel process, and $\bar
u_{r_2}(p_4)$ and $u_{r_1}(p_2)$ are the spinors of the outgoing
$\Sigma^0$ baryon and the initial proton, respectively. The reduced
${\cal A}_i$ read,
\begin{eqnarray}
{\cal A}^{\Lambda(1115)}_s &=& -i g_{KN\Lambda}g_{\pi \Sigma \Lambda}
\frac{{\not \! p_1}+{\not \! p_2} - m_{\Lambda}}{ s - m_{\Lambda}^2} , \\
{\cal A}^{\Lambda(1405)}_s &=& i g_{\Lambda^* {\bar K}N}g_{\Lambda^*
\pi\Sigma} \frac{{\not \! p_1}+{\not \! p_2}+M_{\Lambda^*}}{ s - M_{\Lambda^*}^2+iM_{\Lambda^*}\Gamma_{\Lambda^*}} , \\
{\cal A}_t &=&  - \frac{g_{K^*K\pi} g_{K^*\Sigma N}}{q^2-m^2_{K^*}}
( {\not \! p_1}+{\not \! p_3}-
\frac{m^2_{K}-m^2_{\pi}}{m^2_{K^*}} {\not \! q} \nonumber \\
&& -\frac{\kappa}{m_N}(p_1\cdot p_3-{\not \! p_1}{\not \! p_3}) ) , \\
{\cal A}_u &=&  g_{K\Sigma N} g_{\pi NN} \frac{{\not \! p_2}-{\not
\! p_3}-m_N}{ u - m^2_N}.
\end{eqnarray}
where $q$ is the momentum of exchanged meson $K^*$ in the
$t-$channel, while $s=(p_1+p_2)^2$, is the invariant mass square of
the $K^- p$ system. As we can see, in the tree-level approximation,
only the products like $g_{\Lambda^*{\bar K} N} g_{\Lambda^*
\pi\Sigma}$ enter in the invariant amplitudes. They are determined
by fitting them to the low energy experimental data on the total
cross sections of $K^-p \to \pi^0\Sigma^0$ reaction~\cite{pidata3}
with the usage of the MINUIT fitting program. Besides,
$M_{\Lambda^*}$ and $\Gamma_{\Lambda^*}$ are the mass and total
decay width of the $\Lambda(1405)$ resonance, which we will take the
average values as quoted in the PDG~\cite{pdg2012}.

As we are not dealing with point-like particles, we ought to
introduce the compositeness of the hadrons. This is usually achieved
by including form factors in the finite interaction vertexes. In the
present work, we adopt the following form
factors~\cite{wufq,Mosel,feuster,oh}
\begin{equation}\label{FB}
F(q_{ex}^2,M_{ex})={\Lambda^4\over
\Lambda^4+(q_{ex}^2-M_{ex}^2)^2}\, ,
\end{equation}
for all the channels, where the $q_{ex}$ and $M_{ex}$ are the
4-momenta and the mass of the exchanged hadron, respectively. In
present calculation, the cutoff parameter $\Lambda$ is constrained
between $0.6$ and $1.2$ GeV for all channels.

\section{Numerical results and discussion} \label{Results}

The differential cross section for $K^-p \to \pi^0 \Sigma^0$ at
center of mass (c.m.) frame can be expressed as
\begin{equation}
{d\sigma \over d{\rm cos}\theta}={1\over 32\pi s}{
|\vec{p_3}^{\text{c.m.}}| \over |\vec{p_1}^{\text{c.m.}}|} \left (
{1\over 2}\sum_{r_1,r_2}{|\cal M|}^2 \right ), \label{eq:dcs}
\end{equation}
where $\theta$ denotes the angle of the outgoing $\pi^0$ relative to
beam direction in the $\rm c.m.$ frame, $\vec{p_1}$ and $\vec{p_3}$
are the 3-momentum of the initial $K^-$ and the final $\pi^0$
mesons, respectively, while the total invariant scattering amplitude
$\cal M$ is given by~\footnote{ In phenomenological Lagrangian
approaches, the relative phases between different amplitudes are not
fixed. In general, we should introduce a relative phase between
different amplitudes as a free parameter, since now we have only few
experimental data on the total cross sections, which are not
sensitive to the relative phases, so we take all the relative phases
as zero.},
\begin{equation}
{\cal M}={\cal M}_s + {\cal M}_t+ {\cal M}_u \, .
\end{equation}

Firstly, by including all the contributions from $s-$channel
$\Lambda(1405)$ resonance, $s-$channel $\Lambda(1115)$, $t-$ channel
$K^*$, and $u-$channel proton processes, at a fixed cut off
parameter $\Lambda$, we perform a $\chi^2$ fit (Fit I) to the total
cross section data taken from Ref.~\cite{pidata3}. It is worth
noting that since we want to get the coupling constant
$g_{\Lambda(1405)\bar{K}N}$ from the total cross sections of the
$K^- p \to \pi^0 \Sigma^0$ reaction, so the contributions from other
$\Lambda$ resonance are neglected. In the present work, only the
experimental data very close to the reaction threshold are taken
into account, while those data below the production threshold of
$\Lambda(1520)$, which is the next $\Lambda$ resonance above the
$\Lambda(1405)$, are neglected. Then there are $10$ available data
points totally.

\begin{figure}[htbp]
\begin{center}
\includegraphics[scale=0.45]{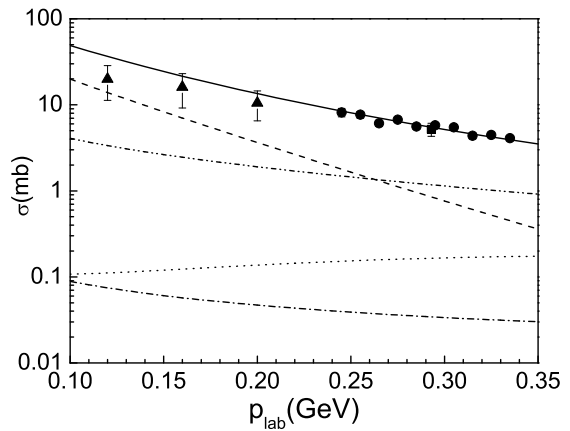}
\caption{$K^- p \to \pi^0 \Sigma^0$ total cross sections compared
with the experimental data from Ref.~\cite{pidata3} (dot),
Ref.~\cite{pidata1} (triangle), and Ref.~\cite{pidata2} (square).
Results have been obtained from the Fit I. The solid line represents
the full results, while the contribution from $s-$channel
$\Lambda(1405)$ resonance, $s-$channel $\Lambda(1115)$, $t-$ channel
$K^*$, and $u-$channel proton processes are shown by the dashed,
dotted, dot-dashed, and dot-dot-dashed lines, respectively.}
\label{tcsall}
\end{center}
\end{figure}

By constraining the value of the cutoff parameter $\Lambda$ from
$0.6$ to $1.2$ GeV, we get the minimal $\chi^2/dof$ is $0.9$ with
$\Lambda = 0.6$ GeV for all the channels, and the fitted parameter
$g_{\Lambda^*\bar{K}N} g_{\Lambda^* \pi \Sigma}$ is $ -0.70 \pm
0.06$. From the value of the $g_{\Lambda^* \pi\Sigma}$, we can get
$|g_{\Lambda^* \bar{K} N}|= 0.77 \pm 0.07$.

\begin{figure}[htbp]
\begin{center}
\includegraphics[scale=0.45]{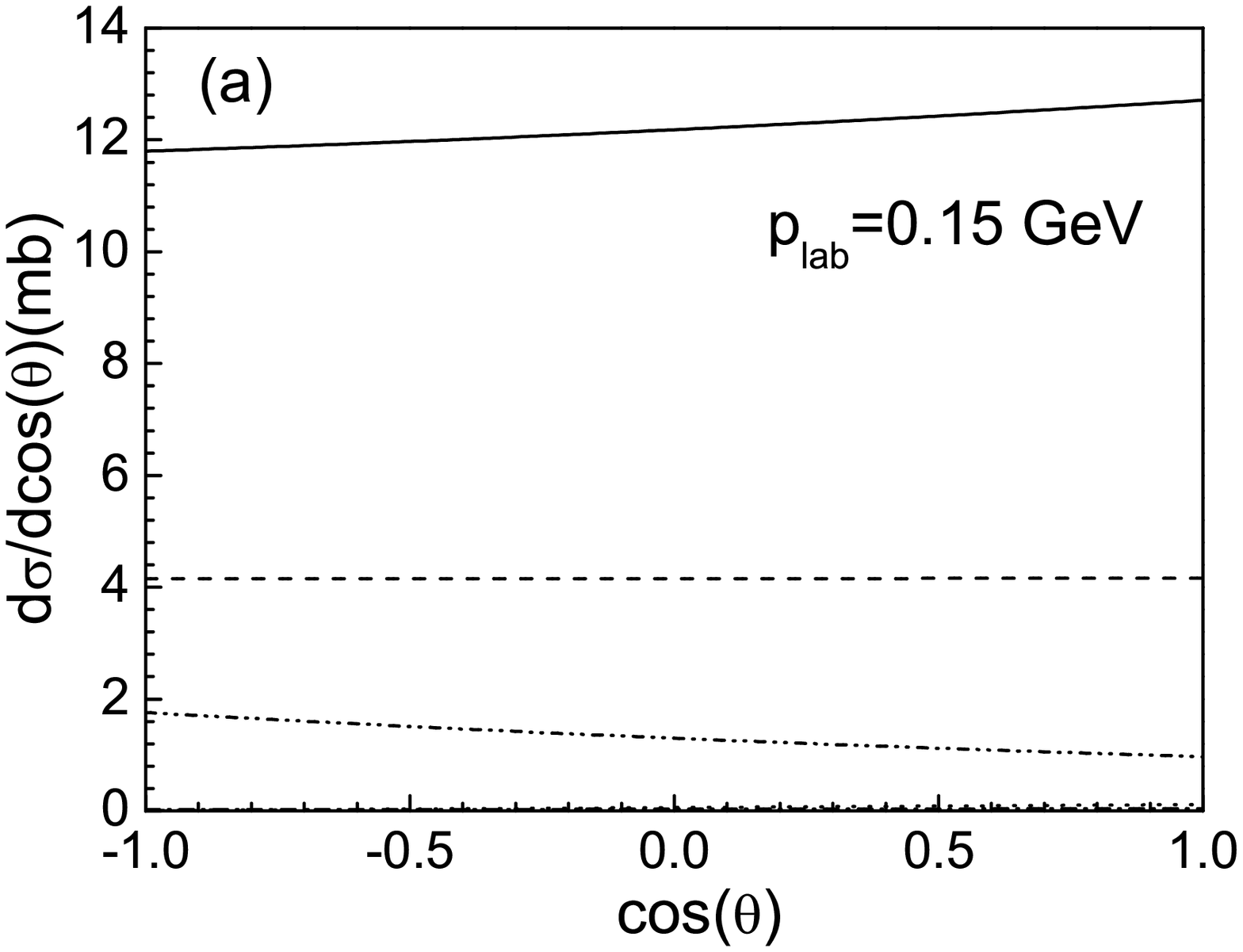}
\includegraphics[scale=0.45]{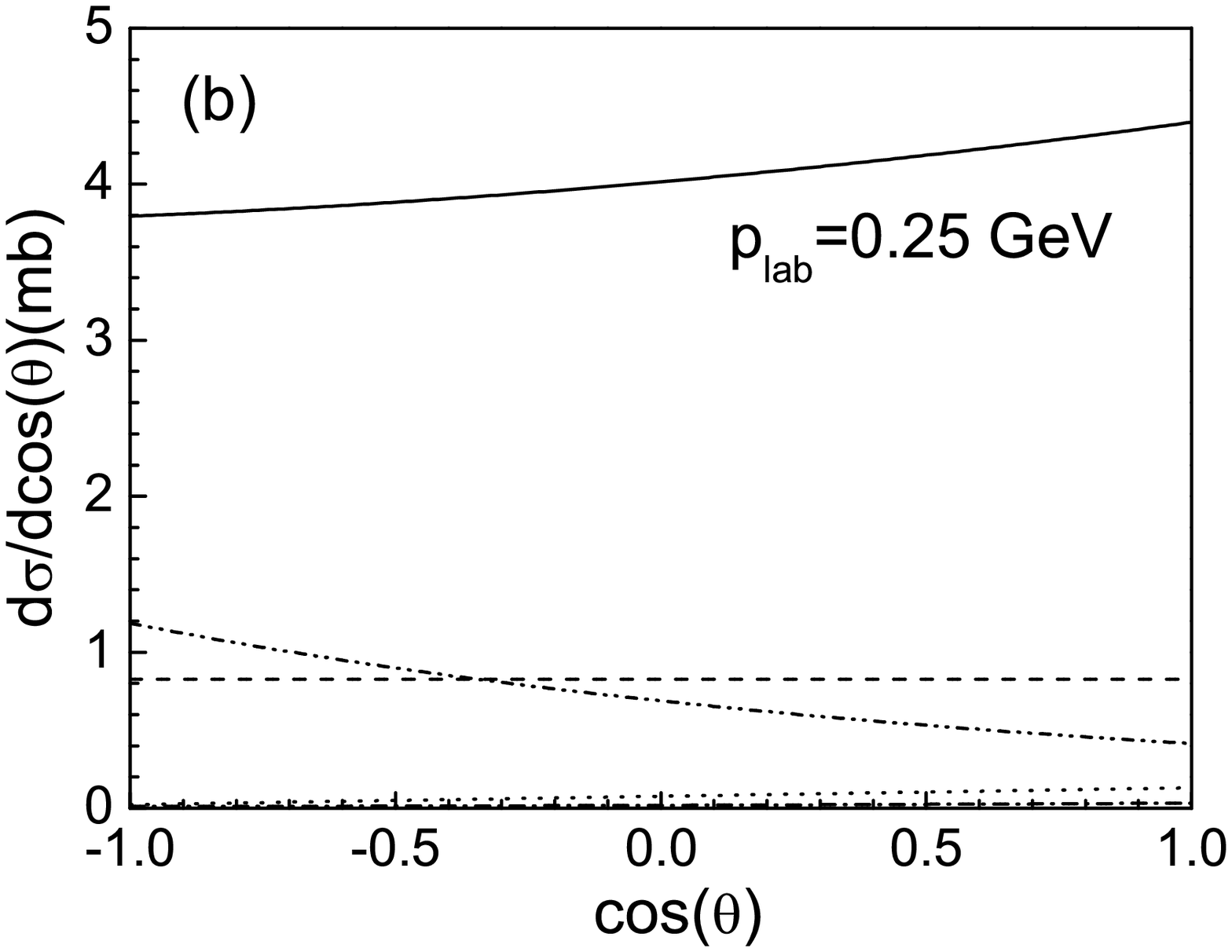}
\includegraphics[scale=0.45]{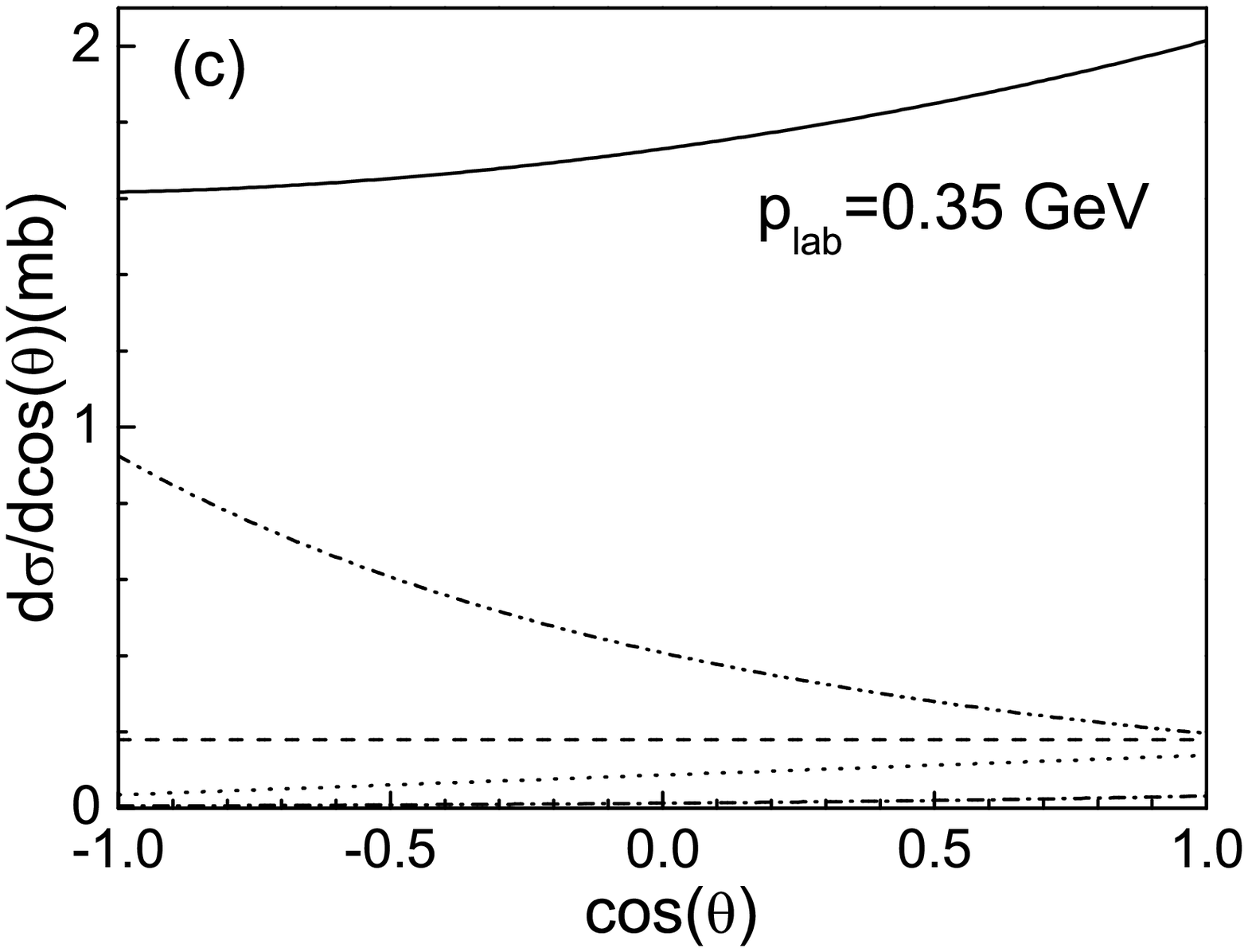}
\caption{$K^- p \to \pi^0 \Sigma^0$ differential cross sections at
different energies. Results have been obtained from the Fit I. The
solid line represents the full results, while the contribution from
$s-$channel $\Lambda(1405)$ resonance, $s-$channel $\Lambda(1115)$,
$t-$ channel $K^*$, and $u-$channel proton processes are shown by
the dashed, dotted, dot-dashed, and dot-dot-dashed lines,
respectively.} \label{dcs}
\end{center}
\end{figure}

The corresponding best fitting results of the Fit I for the total
cross sections are shown in Fig.~\ref{tcsall}, comparing with the
experimental data. We also show, in Fig.~\ref{tcsall}, the
experimental data with larger error from Ref.~\cite{pidata1} and one
data point from Ref.~\cite{pidata2} for comparison. The solid line
represents the full results, while the contributions from
$s-$channel $\Lambda(1405)$ resonance, $s-$channel $\Lambda(1115)$,
$t-$, and $u-$channel diagrams are shown by dashed, dotted, and
dot-dashed lines, respectively. From Fig.~\ref{tcsall}, one can see
that we could describe the near threshold data of $K^- p \to \pi^0
\Sigma^0$ reaction quite well, and the $s-$channel $\Lambda(1405)$
resonance and $u-$channel proton pole give the dominant
contributions, while the $s-$channel $\Lambda(1115)$ process and
$t-$channel $K^*$ exchange give minor contribution. Theoretically,
it is important to find some observables to distinguish the relative
roles of individual mechanisms. In Fig.~\ref{dcs}, the corresponding
theoretical calculation results for the differential cross sections
at p$_{\text {lab}}=0.15$ GeV [Fig.~\ref{dcs} (a)], p$_{\text
{lab}}=0.25$ GeV [Fig.~\ref{dcs} (b)], and p$_{\text {lab}}=0.35$
GeV [Fig.~\ref{dcs} (c)] are shown, which can be tested by future
experiments.

\begin{figure}[htbp]
\begin{center}
\includegraphics[scale=0.45]{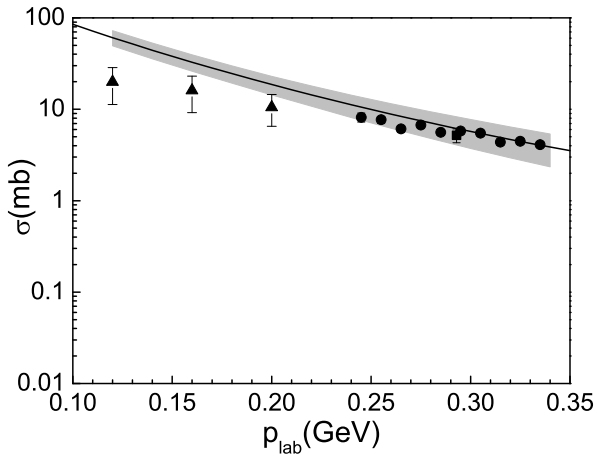}
\caption{As in Fig.~\ref{tcsall}, but for the best fitted results of
the Fit II. We also the error band which is obtained from the
uncertainties of the fitted parameters.} \label{tcs2}
\end{center}
\end{figure}

For the role of $u-$channel proton pole diagram, since there could
be large SU(3) flavor symmetry violation, as summarized in
Ref.~\cite{General:2003sm} (see Table II of this reference), the
value of $g_{KN\Sigma}$ lies in a very wide range. Besides, the
contributions from $s-$channel $\Lambda(1115)$ and $t-$channel
diagram are very small, so, next, we try the fit with considering
the contribution from only $s-$channel $\Lambda(1405)$ resonance
(Fit II). In this case, we have two free parameters which are
$g_{\Lambda^* {\bar K}N}g_{\Lambda^*\pi\Sigma}$ and the cut off
parameter $\Lambda^{\Lambda(1405)}_s$. The fitted results are
$g_{\Lambda^* {\bar K}N}g_{\Lambda^* \pi\Sigma} = 1.36 \pm 0.08$,
which gives $|g_{\Lambda^*\bar{K}N}| = 1.51 \pm 0.10$, and
$\Lambda^{\Lambda(1405)}_s =3.00 \pm 2.62$ GeV, with a large
$\chi^2/dof=1.7$. The corresponding results for the total cross
sections are shown in Fig.~\ref{tcs2} with the solid line. We also
show the 90\% confidence-level band obtained from the statistical
uncertainties of the fitted parameters.~\footnote{We generate pairs
of parameters ($g_{\Lambda^* {\bar K}N}g_{\Lambda^*\pi\Sigma}$ and
$\Lambda^{\Lambda(1405)}_s$) from a two-dimensional correlated
Gaussian distribution with the mean values and standard deviations
obtained from the best $\chi^2$ fitt. For each ($g_{\Lambda^* {\bar
K}N}g_{\Lambda^*\pi\Sigma}$ and $\Lambda^{\Lambda(1405)}_s$) pair,
we calculate the total cross sections. We plot all these results and
throw away the upper 5\% and the lower 5\%, then the band, shown in
Fig.~\ref{tcs2}, is obtained.} The results show that we can also
give a reasonable description for the experimental data by only
including the $s-$channel $\Lambda(1405)$ resonance. However, even
by considering the errors of the theoretical calculation, we still
can not give a reasonable description for the data points from
Ref.~\cite{pidata1}.

The value of $|g_{\Lambda^*\bar{K}N}| = 1.51 \pm 0.10$ is comparable
with the value, $1.84$~\cite{Xie:2010md}, which was obtained from
the separable potential model~\cite{gal}. In contrast to the unitary
chiral theory~\cite{chiraloset}, the separable model produces only a
single $\Lambda(1405)$ pole and this is consistent with our
assumption that we only include one $\Lambda(1405)$ state in the
present calculation for simplicity.

\begin{figure}[htbp]
\begin{center}
\includegraphics[scale=0.45]{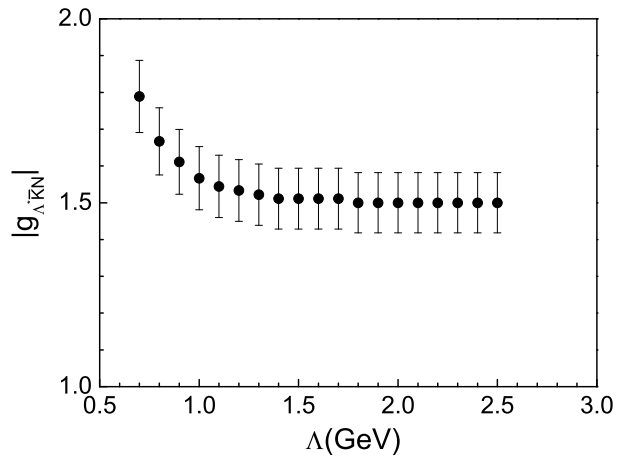}
\caption{Coupling constant $|g_{\Lambda^*\bar{K}N}|$ \emph{versus}
the cut-off parameter $\Lambda$.} \label{gvslam}
\end{center}
\end{figure}

From the fitted parameters of Fit II, we find an unrealistic central
value of $3.00$ GeV for the cutoff $\Lambda^{\Lambda(1405)}_s$
parameter, with a large error ($2.62$ GeV), which indicates that the
$\chi^2$ is rather insensitive to this parameter, so, next we fix
the cut-off at some values, then fit the only one parameter
$g_{\Lambda^* {\bar K}N}g_{\Lambda^* \pi\Sigma}$ to the total cross
sections data, from which we can get the fitted coupling constant
$g_{\Lambda^* {\bar K}N}$ as a function of the cut-off parameter
$\Lambda^{\Lambda(1405)}_s$. The results are shown in
Fig.~\ref{gvslam}. We can see from the figure that the value of
$|g_{\Lambda^* {\bar K}N}|$ is stable at $1.5$ within a very wide
range of the cut-off parameter $\Lambda_s^{\Lambda(1405)}$.

From the values $|g_{\Lambda^* {\bar K}N}| = 1.51 \pm 0.10$ and
$|g_{\Lambda^*\pi \Sigma}| = 0.90 \pm 0.02$, we can easily obtain
the ration $R=|g_{\Lambda^* {\bar K}N}/g_{\Lambda^* \pi \Sigma}| =
1.68 \pm 0.12$~\footnote{If we take the values $0.77$ from Fit I and
$1.51$ from Fit II as the lower and upper limit for the
$g_{\Lambda^*\bar{K} N}$, respectively, then we can get
$|g_{\Lambda^*\bar{K} N}| = 1.14 \pm 0.37$, which gives $R = 1.27
\pm 0.41$.}, which is smaller than those obtained from different
models: $2.19$ obtained by using an algebra-of-currents
approach~\cite{Weil}, and $2.61\pm 1.34$ extracted from the
coupled-channel analysis of the $K^{-}p$ scattering ~\cite{jkk}. We
show these values and also the couplings of $\Lambda^* {\bar K}N$
and $\Lambda^*\pi \Sigma$ in the Table~\ref{tab3} for comparison.
From table~\ref{tab3}, we find that even the values of $R$ are
different, but the values for $|g_{\Lambda^* {\bar K}N}|$ are
similar within the errors.

\begin{table}[htbp]
\caption{Couplings of $\Lambda(1405) {\bar K}N$ and
$\Lambda(1405)\pi \Sigma$. }
\begin{tabular}{c|c|c|c}
\hline\hline   $|g_{\Lambda^* {\bar K}N}|$ & $|g_{\Lambda^* \pi\Sigma}|$ & $R$ & \\
\hline
$1.64$ & $0.75$ & $2.19$ & \cite{Weil} \\
$2.10 \pm 0.71$ & $0.77 \pm 0.30$ & $2.61 \pm 1.34$ & \cite{jkk} \\
$1.51 \pm 0.10$ & $0.90 \pm 0.02$ & $1.68 \pm 0.12$ & This work \\
\hline \hline
\end{tabular}
\label{tab3}
\end{table}

On the other hand, in analysis of the line shapes of the $\Sigma
\pi$ final states in the production reaction $\gamma  p \to K^+ +
(\Sigma \pi)$, it is convenient to parameterize the scattering
amplitude, in the isospin zero sector, as the Breit-Wigner function
$\text{BW}(W)$, which has been used in Ref.~\cite{clasnew},
\begin{eqnarray}
\text {BW} (W)=\frac{1}{W^2-M^2_{\Lambda^*}+ i
M_{\Lambda^*}\Gamma_{\Lambda^*}(W)},
\end{eqnarray}
where $W$ is the invariant mass of the $\Sigma \pi$ system, and
$\Gamma_{\Lambda^*}(W)$ is the energy dependent width that accounts
for all the decay channels, this is because of that in the $\gamma p
\to K^+ + (\Sigma \pi)$ reaction, the $N\bar{K}$ channel opens
within the range of the mass distribution of the $\Sigma \pi$
system.

Considering the $\Lambda^*(1405)\bar{K}N$ coupling, by using the
Flatt\'e prescription~\cite{Flatte:1976xu}, the energy dependent
total decay width for the $\Lambda^*(1405)$ resonance,
is~\footnote{We mention that in the present calculation, since the
$\bar{K}N$ channel is opened, so we just use a constant total decay
width for $\Lambda(1405)$ resonance, in such a way we can also
reduce the number of free parameters. Furthermore, we do not
continue the decay momentum to imaginary values, which means below
the threshold of decay channel, we take the decay momentum as zero.}

\begin{eqnarray}
&& \Gamma_{\Lambda^*}(W) = \frac{3g^2_{\Lambda^*\pi\Sigma}}{4\pi}[E_{\Sigma} + m_{\Sigma}]\frac{|\vec{p_{\Sigma}}|}{W} + \nonumber \\
&& \frac{g^2_{\Lambda^* \bar{K} N}}{2\pi} [ E_N + m_N ]
\frac{|\vec{p_N}|}{W} \theta (W - m_{\bar{K}} -m_N), \ \ \ \ \
\label{gamrs}
\end{eqnarray}
with,
\begin{eqnarray}
E_{\Sigma /N} &=& \frac{W^2 + m^2_{\Sigma /N} - m^2_{\pi /\bar{K}}}{2 W}, \\
|\vec{p_{\Sigma /N}}| &=& \sqrt{E^2_{\Sigma /N} - m^2_{\Sigma /N}}.
\end{eqnarray}

\begin{figure}[htbp]
\begin{center}
\includegraphics[scale=0.45]{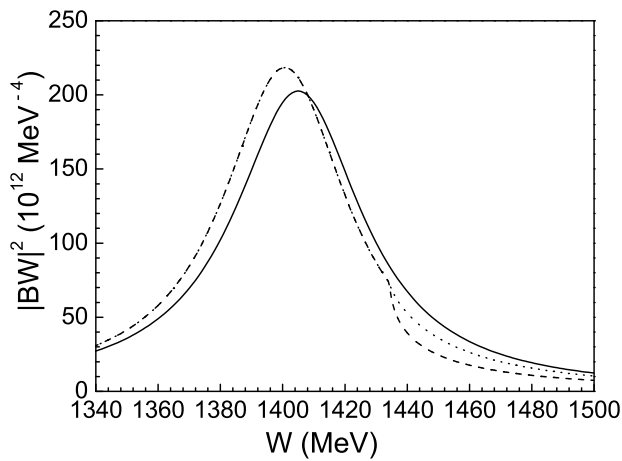}
\caption{The module square of the Breit-Wigner function for
$\Lambda^*(1405)$ vs $W$ with a constant total decay width (solid
line) and an energy dependent width with the form in
Eq.~(\ref{gamrs}) with $g_{\Lambda^*\bar{K}N} = 1.51$ (dashed line)
and $g_{\Lambda^*\bar{K}N} = 0.77$ (dotted line).} \label{bw}
\end{center}
\end{figure}

In Fig.~\ref{bw}, we show the results for the module square of
$\text {BW} (W)$ as a function of $W$. The solid line stands for the
results obtained with a constant total decay width ($50$ MeV), while
the dashed and dotted line are obtained with $g_{\Lambda^*\bar{K}N}
= 1.51$ and $g_{\Lambda^*\bar{K}N} = 0.77$ by using the energy
dependent width with the form in Eq.~(\ref{gamrs}), respectively.
From the results, we find that the Breit-Wigner mass will be pushed
down if we use the energy dependent width, and there is a clear drop
in the module squared distribution at the $\bar{K}N$ threshold with
a larger value $g_{\Lambda(1405)\bar{K}N} = 1.51$. However, the
smaller value $g_{\Lambda(1405)\bar{K}N} = 0.77$ could not give a
clear drop for the module square distribution at the $\bar{K}N$
threshold. The very recent experimental results measured by the CLAS
Collaboration in Ref.~\cite{clasnew} show that there is a sharp drop
of the $\Sigma \pi$ mass distributions, and this could be reproduced
by using the above $\text {BW}(W)$ formalism with a large coupling
constant $g_{\Lambda(1405)\bar{K}N}$.

\section{Summary}

In this work, the value of $\Lambda(1405){\bar K}N$ coupling
constant $g_{\Lambda(1405)\bar{K}N}$ is obtained by fitting it to
the low energy experimental data of the $K^-p \to \pi^0\Sigma^0$
reaction. On the basis of an effective Lagrangian approach, we show
that the value of $\Lambda(1405){\bar K}N$ coupling constant
$|g_{\Lambda(1405)\bar{K}N}| = 0.77 \pm 0.07 $ can be extracted from
the available low energy experimental data of the total cross
section of the $K^-p \to \pi^0\Sigma^0$ reaction by including the
$s-$channel $\Lambda(1405)$ resonance, $s-$channel $\Lambda(1115)$
process, $t-$channel $K^*$ and $u-$channel proton pole diagrams. On
the other hand, by including only the contribution from the
$s-$channel $\Lambda(1405)$ resonance, we get
$|g_{\Lambda(1405)\bar{K}N}| = 1.51 \pm 0.10$, which is supported by
the previous calculations~\cite{Weil,jkk} as shown in
Table~\ref{tab3}.

Due to the violation of SU(3) flavor symmetry, the contribution from
$u-$channel could be small, and we show that the differential cross
sections shown in Fig.~\ref{dcs} could help us clarifying whether
the $u-$channel contribution is important or not.

We also calculate the module square of Briet-Wigner function for the
$\Lambda(1405)$ resonance with a energy dependent total decay width
with the form as the Flatt\'e prescription~\cite{Flatte:1976xu}. The
results show that the Breit-Wigner mass of $\Lambda(1405)$ resonance
will be pushed down, and there is a clear drop in the module square
distribution at the $\bar{K}N$ threshold if we take a larger value
$g_{\Lambda(1405)\bar{K}N} = 1.51$. The clear drop could explain the
sharp drop that was found in the $\Sigma \pi$ mass distribution in
the recent experimental measurements~\cite{clasnew} by the CLAS
Collaboration.

Finally, we would like to stress that the coupling constant
$g_{\Lambda(1405)\bar{K}N}$ is important for studying the
$\Lambda(1405)$ resonance in the $\gamma p \to K^+ \Lambda(1405) \to
K^+(\pi \Sigma)$ reaction and also in the $pp \to pK^+\Lambda(1405)
\to pK^+(\pi\Sigma)$ reaction by using the effective Lagrangian
approach~\cite{Xie:2010md,l1520data,Xie:2013db}. More and accurate
data for these reactions can be used to improve our knowledge on the
structure and properties of $\Lambda(1405)$ state, which are, at
present, still controversial.

\section*{Acknowledgments}

We would like to thank Xian-Hui Zhong, Jun He, and Xu Cao for useful
discussions. This work is partly supported by the National Natural
Science Foundation of China under grants  11105126, 10905046 and
11205164.

\end{document}